\documentclass{article}

\usepackage{amssymb}          
\usepackage{graphicx}         
\usepackage[latin1]{inputenc} 
\usepackage{amsmath}          

\usepackage{amsfonts}

\usepackage{latexsym}
\usepackage{graphicx}
\usepackage{color}

\newtheorem{proposition}{Proposition}[section]
\newtheorem{definition}{Definition}[section]
\newtheorem{theorem}{Theorem}[section]
\newtheorem{corollary}{Corollary}[section]
\newtheorem{lemma}{Lemma}[section]

\newcommand{\proof}{\noindent {\em Proof.} }
\newcommand{\qed}{\hfill $\Box$ \vskip 2ex}

\newcommand{\Hi}{\mathcal{H}}

\newcommand{\C}{\mathbb{C}}

\newcommand{\R}{\mathbb{R}}

\newcommand{\I}{\mathbb{I}}

\newcommand{\ddt}{\frac{d}{d t}}
\newcommand{\um}{\frac{1}{2}}

\newcommand{\Span}{\textrm{span}}

\newcommand{\tr}{\textrm{tr}}
\newcommand{\trace}{\textrm{tr}}

\def\smallfrac#1#2{{\textstyle\frac{#1}{#2}}}

\newcommand{\beq}{\begin{equation}}
\newcommand{\eeq}{\end{equation}}
\newcommand{\beqa}{\begin{eqnarray}}
\newcommand{\eeqa}{\end{eqnarray}}
\newcommand{\beqan}{\begin{eqnarray*}}
\newcommand{\eeqan}{\end{eqnarray*}}

\newcommand{\setdist}[2]{{\mathfrak d}(#1,#2)}

\title{Stabilization of Stochastic Quantum Dynamics\\ via Open and Closed Loop Control \thanks{Partially supported by the QUINTET and QFUTURE strategic projects of the Dept. of  Information Engineering and University of Padua.}}
\author{Francesco Ticozzi\thanks{F. Ticozzi is with Dipartimento di Ingegneria
dell'Informazione, Universit\`a di Padova, via Gradenigo 6/B,
35131 Padova, Italy ({\tt ticozzi@dei.unipd.it}).}, 
Kazunori Nishio\thanks{K. Nishio is with the Department of Mechanical and Environmental Informatics, Tokyo Institute of Technology, 2-12-1-W8-1, O-okayama, Meguro-ku, Tokyo, Japan ({\tt knishio.cyb@gmail.com})}, and Claudio Altafini\thanks{C. Altafini is with SISSA, via
Bonomea 265, 34136 Trieste, Italy ({\tt altafini@sissa.it}).}
}
\date{\today}

\begin{document}

\maketitle
\begin{abstract}

In this paper we investigate parametrization-free solutions of the problem of quantum pure state preparation and subspace stabilization by means of Hamiltonian control, continuous measurement and quantum feedback, in the presence of a Markovian environment.
In particular, we show that whenever suitable dissipative effects are induced either by the unmonitored environment or by  non Hermitian measurements, there is no need for feedback control to accomplish the task.
Constructive necessary and sufficient conditions on the form of the open-loop controller can be provided in this case. When open-loop control is not sufficient, filtering-based feedback control laws steering the evolution towards a target pure state are provided, which generalize those available in the literature.

\end{abstract}


\section{Introduction}

In the ever growing spectrum of applications of control-theoretic methods to quantum mechanical systems, and in particular in the control of molecular dynamics \cite{Peirce1988,Tannor1985,brumer92}, in cooling nano-electromechanical devices \cite{nanomechanical1}, in controlling optical systems \cite{feedbackcooling,orozco1} and in manipulating a variety of physical supports for quantum information processing \cite{nielsen2000,wiseman-book}, state preparation problems maintain a central role.
In particular, pure states, the maximal information states for quantum systems, are sought, e.g. as initial states for enacting controlled energy transitions, as states of minimal energy for cooling procedures, and, especially entangled ones, as fundamental resources for quantum information processing.
When the system state is not pure, or not precisely known, the ``classical'' Lie-algebraic or optimal, open-loop, coherent control design methods are not sufficient to attain purification, and some interaction with an external system, environment or measurement apparatus is needed in order to complete the control task.

We here consider the problem of global asymptotic stabilization of pure states, and subspaces, of finite-dimensional systems interacting with a Markovian environment and undergoing continuous measurement processes.
The most general description of these systems is provided by the so-called Stochastic Master Equation (SME), the quantum equivalent of a Kushner-Stratonovich equation for the probability density in the classical case \cite{doherty2000,mirrahimi2007}. This can be obtained from a suitable 
system/field interaction model by means of non-commutative filtering theory \cite{belavkin1992,bouten2007}. Quantum subspace stabilization is of key interest in quantum information encoding and error correction, and can often be treated with slight generalizations of the same methods \cite{ticozzi2008}.

From a control-theoretic viewpoint, the problem is doubtless challenging. The SME is a nonlinear, control-affine, matrix Stochastic Differential Equation (SDE) with state in the compact set of positive-semidefinite, unit-trace Hermitian matrices. In addition to this, some intrinsic symmetries in the action of noise prevent standard design methods to work.
Different approaches have been used: we refer the reader to the survey papers \cite{Mabuchi2005,petersen-survey} and the book \cite{wiseman-book} for a comprehensive review.  In particular, stabilizing feedback laws based on the real-time estimate of the system state (hence often called bayesian, or filtering-based techniques) have been presented in \cite{doherty1999, handel2005,yamamoto2007}. In particular, the approach of \cite{handel2005} is based on a (convex) numerical Lyapunov design, viable for low-dimensional systems, since the numerical design procedure suffers from scalability problems, being based on explicit parametrizations whose sixe grow quadratically with the dimension of the system. The filtering-based approach has also been shown to be robust with respect to small delays in the control loop \cite{yamamoto-delay}, and to initialization of the filter \cite{vanhandelstability}.   A more general theoretical framework to solve the problem has been presented in \cite{mirrahimi2007}: here the need for numerical methods is bypassed by resorting to switching controllers. However, in order to apply this approach to systems that differ from the ones considered in \cite{mirrahimi2007}, one would have to tailor the solution to the problem under consideration by adapting part of the results to each specific model.

In order to find a constructive and flexible approach to control design, we here review this class of control problems bringing in the picture two new elements: open-loop, time-invariant Hamiltonian control, and Markovian environments.
Assuming we can engineer open-loop, time-independent Hamiltonians on the system, we provide necessary and sufficient conditions for pure state and subspace stochastic stabilizability (in probability) for a large class of systems, exploiting either the dissipation effects induced by the measurement or the noise processes but without requiring feedback control.
It is worth observing that in practical applications the use of real-time feedback is an extremely challenging task which entails the computation-intensive real-time integration of the SME, while open-loop, time-independent solutions appear to be a more viable option in many practical situations \cite{ticozzi-NV}. 

Other ways of substituting real-time feedback with open-loop control have also been sought in \cite{wiseman-openloop}, with some key differences: the open loop control is ``impulsive'', i.e. consists in a sequence of unitary rotations at discrete time intervals, while the quantum filter acts continuously on the whole control interval; the tasks considered are those of rapid purification, where the target state is not decided a priori, but it is only required to be pure, or as pure as possible, as well as rapid measurements.

When no suitable dissipation effect (neither monitored nor unmonitored) is available, we show that open-loop control is not enough to achieve the desired stabilization. 
In this case we can resort to a feedback method, and we illustrate how a time-dependent Hamiltonian, driven by a suitable feedback control law can bridge the gap, ensuring stabilization at least in expectation of a target pure state. The design of the feedback law we pursue follows \cite{mirrahimi2007}, with some differences that are worth remarking: (i) it takes into account, and exploits, the effect of the open-loop Hamiltonian and the environment; (ii) it is parametrization free, and the method directly applies to a fairly wide class of models; (iii) the assumption we make are such that the role of the feedback part in the control strategy is ``minimal'', i.e. is used only to enable state transitions that the open-loop control cannot produce. Further implications of these facts will be discussed in the conclusions.

The paper is structured as follows: in Section \ref{models} we recall
the class of models we will consider, along with some of their basic properties. Section \ref{secinvariance} is devoted to characterize invariant subspaces, and subspaces that cannot be such, for the SME. We do so by resorting to the properties of an associated deterministic system: a key role is played by the {\em support theorem} \cite{mirrahimi2007, kunita-flow,kunita-supp}, which is recalled in Appendix \ref{stochstab} together with some results of stochastic stability that will be needed in the following sections.
Building on these results, Section \ref{EAS} provides necessary and sufficient condition for the open-loop, {\em environment-assisted} stabilization of the SME. When these conditions are not matched, we illustrate an effective strategy including feedback in Section \ref{Sec:FAS}.


\section{Models for quantum dynamics with noise and measurement processes}\label{models}
We need to recall some basics of quantum mechanics, quantum
filtering and stochastic stability theory we will use later on.
For an excellent introductory exposition to the statistical description
of quantum systems see e.g. \cite{maassen-qp}. More
details can be found in e.g. \cite{meyer,parthasarathy} and
references therein.
For the theory of stochastic stability, main references are \cite{Arnold1,khasminskiy1980,kushner1967}: we recall the results of interest in Appendix \ref{stochstab}.

A finite-dimensional quantum system ${\cal Q},$ with Hilbert space
${\cal H}$ over the complex field $\C,$ is considered. 
Let ${\mathfrak B}({\cal H})$ represent the set
of linear operators on ${\cal H}$, with ${\mathfrak H}({\cal
H})$ denoting the real subspace of Hermitian operators, and $\I$ being
the identity operator. The adjoint of $A\in{\mathfrak B}({\cal H})$
denoted by $A^\dagger$. Our knowledge of the state of ${\cal Q}$ is
condensed in a density operator $\rho$ on ${\cal H}$, with $\rho\ge 0$
and $\tr(\rho)=1$. Density operators form a convex set, denoted ${\mathfrak D}({\cal H})\subset{\mathfrak H}({\cal
H})$.  Observables are represented by Hermitian operators in ${\mathfrak
H}({\cal H})$.

\subsection{Quantum filtering equation}

The most general and rigorous method to derive the state dynamics for a given system, while
this is subjected to continuous observations, relies on quantum probability theory,
and in particular quantum filtering theory \cite{belavkin1985,belavkin1992,
bouten2007}. The result is a stochastic differential
equation for the estimated state conditioned on the measurements, driven by the noise process deriving from the 
measurement process. We recall in the following the form of dynamics including only homodyne-detection measurements
\cite{wiseman1994,barchielli2009}, and hence driven by continuos gaussian 
white noise. While for the sake of simplicity we consider only one measurement channel at a time, the results could be extended to multiple commuting measurement processes (see also \cite{wiseman-MIMO}). 

Let $ ( \Omega, {\cal E}, P ) $ a (classical) probability space and $ \{
W_t , t\in \mathbb{R}^+ \} $ a standard $ \mathbb{R}$-valued Wiener
process defined on this space. The homodyne-detection measurement record
$y_t$ can be represented as: \beq \label{photocurrent} dy_t=\sqrt{\eta}\frac{1}{2}\tr(\rho_t
(L_0+L_0^\dag))dt+dW_t, \eeq where $L_0$ is the linear operator associated to the measurement process and
$0\leq\eta\leq 1$ represents the efficiency of the measurement. We denote
by $\mathcal{E}_t$ the filtration associated to $ \{ W_t , t\in
\mathbb{R}^+ \} $.

The associated
evolution of the state $\rho_t\in{\mathfrak D}(\Hi)$ conditioned on the
measurement record (\ref{photocurrent}) is the quantum filtering or
Stochastic Master Equation (SME) \`a la It\^o:
\begin{equation}
d\rho_t=\left\{{\cal F}(H,\rho_t)+\sum_{k=0}^r{\cal D}(L_k,\rho_t)
\right\} dt + {\cal G}(L_0,\rho_t) d W_t, \label{eq:SME-Ito1}\eeq
where
\beq\begin{split}
&\hspace{-9mm}{\cal F}(H,\rho):=-i[H,\rho],\\
&\hspace{-9mm}{\cal D}(L,\rho):=L\rho L^\dag-\frac{1}{2}(L^\dag L\rho+\rho L^\dag L),\\
&\hspace{-9mm}{\cal G}(L,\rho):=\sqrt{\eta}
(L\rho+\rho L^\dag-\tr( (L+L^\dag) \rho) \rho).
\end{split}
\end{equation}
The term $ {\cal F} $ is associated to the Hamiltonian of the system, which is given by a
drift and a (bilinear) control part $ H = H_o + u H_f $, while $ {\cal D}(L_0,\rho) $
and $ {\cal G}(L_0,\rho) $ are the drift and diffusion parts introduced by the weak
measurement of the operator $L_0\in{\mathfrak
B}(\Hi_Q)$.  
The additional drift terms $ {\cal D}(L_k,\rho),$ determined by the {\em noise operators}
$L_k\in{\mathfrak B}(\Hi_Q),$ $k=1,\ldots,n,$ account for the non-unitary dynamics induced by interactions with an (unobservable) environment. 
%
The solution $\{\rho_t\}$ uniquely exists and is adapted to the filtration
$\mathcal{E}_t$ and ${\mathfrak D}(\Hi)$-invariant by construction,
see \cite{handel2005,belavkin1992}.  Considering \eqref{photocurrent}
and \eqref{eq:SME-Ito1} together, one can recognize the basic structure
of a {\em Kalman-Bucy} filter. Other
correspondences and differences with the classical setting have been highlighted in 
e.g. \cite{doherty2000,mirrahimi2007}.

\subsection{Quantum dynamical semigroups and their properties}

The drift part of \eqref{eq:SME-Ito1} is linear, and the associated linear system, which is time-invariant if the control $u$ is constant, plays the role of the Kolmogorov's forward equation (or Fokker-Plank equation) associated to the SME \eqref{eq:SME-Ito1}:
\begin{equation}
\label{eq:Lindblad}
\ddt{\rho}(t)={\cal L}(\rho(t))={\cal F}[H,\rho(t)]
+\sum_{k=0}^r {\cal D}(L_k,\rho(t)).
\end{equation}
$\cal L$ is the generator of the Markov semigroup associated to \eqref{eq:SME-Ito1}, i.e. a Quantum Dynamical Semigroup (QDS) generator in the language of statistical quantum mechanics
\cite{lindblad1976,gorini1978}. The resulting evolution is a continuous, one-parameter semigroup of Trace-Preserving, Completely-Positive (TPCP) maps
$\{{\cal T}_t(\cdot)\}_{t\ge 0}$.

In the study of open quantum systems for quantum thermodynamics \cite{alicki-lendi,petruccione}, quantum control \cite{wiseman-book,altafini-open,shabani-lidar, ticozzi2008,ticozzi-NV},
and quantum information \cite{alicki-lendi,petruccione,nielsen2000}, the dynamical properties of QDS's have been thoroughly investigated.
In order to develop a system-theoretic analysis of the system, essentially a theory of controlled invariants \cite{marro-invariants}, a linear algebraic representation and its block decomposition turned out to be insightful \cite{ticozzi2008,ticozzi2009}. We here recall some results on the stability of a subspace $\Hi_S$ of $\Hi$. Consider a decomposition:
\begin{equation}
\Hi=\Hi_S\oplus\Hi_R.
\end{equation}

Let $s=\dim(\Hi_S)$, and let
$\{|\phi_j^S\rangle\}_{j=1}^s$, $\{|\phi_l^R\rangle\}_{l=1}^{n-s}$ be
orthonormal bases for $\Hi_S$, $\Hi_R$, respectively. Then, the
natural basis for $\Hi$,
$\{|\varphi_m\rangle\}=\{|\phi_j^S\rangle\}_{j=1}^s\cup\{|\phi_l^R\rangle\}_{l=1}^{n-s}$,
induces a block structure for matrices acting on $\Hi$:
\begin{equation}
X=
\left(\begin{array}{c|c}
X_S&X_P\\
\hline
X_Q&X_R
\end{array}\right).
\end{equation}
We denote by $\Pi_\bullet$ the projector onto $\Hi_\bullet$. In the block
structure, for instance, $\Pi_S$ is represented in the following form
\[
\Pi_S=
\left(\begin{array}{c|c}
\I_S&0\\
\hline
0&0
\end{array}\right).
\]

The system  with state $\rho\in{\mathfrak D}(\Hi)$ is
initialized in $\Hi_S$ with state $\rho'\in{\mathfrak D}(\Hi_S)$ if $\rho_S=\rho',$ $\rho_P=0$ , $\rho_R=0$. Denote by
${\mathfrak I}_S(\Hi)$ the set of states that satisfies the above
condition for some $\rho'$. With a slight abuse of terminology we say that $\Hi_S$ is an invariant subspace for the dynamics \eqref{eq:Lindblad}
if ${\mathfrak I}_S(\Hi)$ is an invariant subset of
${\mathfrak D}(\Hi)$.
Necessary and sufficient conditions for the blocks of $H$ and $L_k$
to ensure invariance are given by the following proposition \cite{ticozzi2008}.
\begin{proposition}\label{propinvariance}
Assume that $\Hi=\Hi_S\oplus\Hi_R$, and let $H$, $\{L_k\}$ be the
Hamiltonians and the error generators in
(\ref{eq:Lindblad}). Then, $\Hi_S$ is invariant {\em iff}
$\forall k$
\begin{equation}
\begin{split}
&L_k=\left(\begin{array}{c|c}
L_{k,S}&L_{k,P}\\
\hline
0&L_{k,R}
\end{array}\right),\\
&iH_P-\frac{1}{2}\sum_k L_{k,S}^\dagger L_{k,P}=0.
\end{split}
\end{equation}
\end{proposition}
With the same terminology as above, we say that $\Hi_S$ is an ``attractive'' subspace  with respect to a family of TPCP maps $\{{\cal
T}_t\}_{t\ge 0}$ if ${\mathfrak I}_S(\Hi)$ is an attractive set, that is if \begin{equation}
\forall\rho\in{\mathfrak D}(\Hi),\quad\lim_{t\to\infty}\left\|
{\cal T}_t(\rho)-
\left(\begin{array}{c|c}
\rho_S(t)&0\\
\hline
0&0
\end{array}\right)\right\|=0.
\end{equation}
A subspace which is invariant and attractive is called Globally Asymptotically Stable (GAS).
The following theorem characterizes the generators that allow
for open-loop Hamiltonian stabilization of a given subspace \cite{ticozzi2009}.
\begin{theorem}
\label{theorem:ticozzi2009}
Let $\Hi=\Hi_S\oplus\Hi_R$, with $\Hi_S$ invariant for \eqref{eq:Lindblad}. Assume that we can apply arbitrary, time-independent control
Hamiltonians. Then $\Hi_S$ can be made GAS {\em iff}
$\Hi_R$ is not invariant.
\end{theorem}
Hence, by applying Proposition \ref{propinvariance} to $\Hi_R,$ we explicitly see that $\Hi_S$ can be made GAS if, in addition to
the condition $L_{k,Q}=0\;\forall k$, implying the invariance of ${\cal S}$, 
we have that $L_{k,P}\neq0$ for some $k$.
Furthermore, following the proof of Theorem \ref{theorem:ticozzi2009}, a stabilizing Hamiltonian that satisfies the invariance condition can be constructed explicitly \cite{ticozzi2009}.

\section{Invariant and attractive subspaces for the SME} \label{secinvariance}



In order to connect the invariance properties of subspaces of the ME to those of the SME, we will study the properties of a set of linear deterministic differential equations and invoke the {\em support theorem}. This section closely follows the ideas of Section 3 of \cite{mirrahimi2007}, and in particular the support theorem which is recalled in the Appendix (Theorem~\ref{support}). 


In order to prove invariance of ${\mathfrak D}(\Hi)$ and other useful results, in \cite{mirrahimi2007} a linear diffusion, which plays the role of the Zakai equation in the quantum setting, has been used:
\begin{equation}\begin{split}
\label{eq:zakai}
  d\tilde\rho_t=&{\cal F}(H_t,\tilde\rho_t)\,dt
   +\sum_k{\cal D}(L_k,\tilde\rho_t)\,dt\\
   &+\sqrt{\eta}\,(L_0\tilde\rho_t+\tilde\rho_tL_0^*)\,dy_t.
   \end{split}
\end{equation}
This equation preserves nonnegativity of the (pseudo) state $\tilde\rho,$ but does not preserve its trace. However, it is easier to prove the existence of its solution as it obeys global Lipischitz conditions, and the unique strong solution of \eqref{eq:SME-Ito1} can be found from solutions of \eqref{eq:zakai} as $\rho_t=\tilde\rho_t/\trace{\tilde\rho_t}.$ Let us:

\noindent (i) rewrite the initial state in spectral form \[\rho_0=\sum_jp_jv_t^jv_t^{j\dag}=\sum_jp_j\tilde\rho_0^j;\] 

\noindent(ii) introduce a new Wiener process $W_t'$, independent of $y_t,$ and define $\tilde W_t=\sqrt{\eta}y_t+\sqrt{1-\eta}W'_t$. 
We are now in the position to construct a process that is equivalent in law to $\tilde\rho_t$ from the solution of the set of equations:
\beq \label{eq:SME-Ito3}\begin{split}d\tilde\rho^j_t=&{\cal F}(H_t,\tilde\rho^j_t)\,dt
   +\sum_k{\cal D}(L_k,\tilde\rho^j_t)\,dt\\
   &+\sqrt{\eta}\,(L_0\tilde\rho^j_t+\tilde\rho^j_tL_0^\dag)\,d\tilde W_t,
\end{split}\eeq 
and then defining $\bar\rho_t=\mathbb{E}(\sum_jp_j\rho^j_t|{\cal F}^Y_t),$
where ${\cal F}^Y_t$ is the filtration associated to the measurement process $Y_t$.
By Ito's rule one can verify that the solutions $v^j_t$ of linear vector equations in Ito form:
\beq\label{eq:linito} d v^j_t = \left(-i H_t -\um\sum_kL_k^\dag L_k\right)v^j_t\, dt+L_0v^j_td\tilde W_t,\eeq
satisfy $v^j_tv^{j\dag}_t=\tilde\rho^j_t,$ solution of \eqref{eq:SME-Ito3}.
Adding the Wong-Zakai correction\footnote{ Consider a random process $X_t$ on $\C^n,$ two functions $A,B:\C^n\rightarrow\C^n$ ($C^1$ and $C^2,$ respectively), and denote by $X_i,A_i,B_i$ their $i$-th components. An Ito SDE of the form $dX_t=A(X_t)dt+B(X_t)dW_t,$ can be transformed (up to some technical conditions) to Stratonovich form $dX_t=A_S(X_t)dt+B(X_t)\circ dW_t$ 
with corrected drift \cite{ikeda-book} \[A_{S,i}(X)=A_i(X)+\frac{1}{2}\sum_j\frac{\partial B_{i}(X)}{\partial X_j}B_{j}(X).\]} to the drift, we obtain the equivalent Stratonovich form:
\beq \label{eq:linstrat} d v^j_t = \left(-i H_t -\um\sum_kL_k^\dag L_k-\um L_0^2\right)v^j_t\, dt+L_0v^j_t\circ d\tilde W_t.\eeq
Notice that $\tilde W_t=\sqrt{\eta}Y_t+\sqrt{1-\eta}W'_t$ is a process with drift. By Girsanov's theorem we can find another measure $\mathbb{Q}$ that is equivalent to $\mathbb{P}$ and for which $\tilde W_t$ is a Wiener process. 
The last equation, with respect to the measure $\mathbb{Q},$ satisfies all the hypothesis of Theorem \ref{support}. We shall now use it to connect a.s. invariance of subspaces for the SME to their invariance for the ME.


\begin{theorem}\label{propiffinvariance} $\Hi_S$ 
is an invariant subspace for the ME \eqref{eq:Lindblad} if and only if ${\mathfrak I}_S(\Hi)$ is invariant for the SME a.s.
\end{theorem}
\proof Recall from Proposition \ref{propinvariance} that $\Hi_S$ is invariant for \eqref{eq:Lindblad} {\em iff} $L_{k,Q}=0,\,\forall k$ and $-iH_P-\um\sum_k L_{k,S}L_{k,P}^\dag=0.$
Taking into account these conditions in the deterministic differential equation
\beq\label{eq:lindet} \ddt v_t = \left(-i H_t -\um\sum_kL_k^\dag L_k-\um L_0^2\right)v_t+u_tL_0v_t,\eeq we get that both the ``state'' and the ``input'' matrices have an upper-triangular block structure induced by the decomposition $\Hi=\Hi_S\oplus\Hi_R$:
\beq\label{eq:lindet1} \ddt v_t = \left[
\begin{array}{c|c} * & * \\ \hline 0 & * \end{array}\right]v_t+u_t\left[
\begin{array}{c|c} * & *\\ \hline 0 & * \end{array}\right]v_t.\eeq
Hence, for any control input and initial vector $v_0\in\Hi_S,$ it will be $v_t\in\Hi_S$ for any $t>0$.  
Call ${\cal W}_{S,v_0}$ the set of continuous path $v$ on $t\in[0,\infty]$ starting at $v_0$ such that $\Pi_Rv_t=0,\,t\geq 0.$ Since ${\cal W}_{S,v_0}$ is closed in the topology of uniform convergence on compact sets 
\footnote{We can decompose every trajectory $v(t)=\Pi_Sv(t)+\Pi_Rv(t)=v_S(t)+v_R(t),$ and $\|v(t)\|^2=\|v_S(t)\|^2+\|v_R(t)\|^2.$ If a sequence of continuous $\{v^k\}$ converges uniformly on compact sets to $v$, $v$ is continuous and we have for every compact ${\cal K}\subset\R^+$ \[\lim_{k\rightarrow \infty}\sup_{t\in{\cal K}}\|v^k(t)-v(t)\|=0=\lim_{k\rightarrow \infty}\sup_{t\in{\cal K}}\|v^k(t)-v(t)\|^2.\] If $\{v^k\}\in{\cal W}_{S,v_0}$ for all $k$, then $\|v^k_R(t)\|=0$ for all $t,$ and $\|v^k(t)-v(t)\|^2=\|v_S^k-v_S\|^2+\|v_R\|^2.$ In order for the limit above to be zero, it must be $\sup_{t\in{\cal K}}\|v_R(t)\|^2=0$ for every ${\cal K}$. Since we can cover $\R^+$ with compact sets, we have that $v_R(t)=0$ for all $t\geq 0$. Hence each limit $v\in{\cal W}_{S,v_0},$ which is thus closed.}
, from \eqref{eq:support_sub} we have ${\cal S}_{v_0}\subseteq {\cal W}_{S,v_0}.$ By applying the support theorem~\ref{support}, we also have that $1=\mathbb{Q}(\{\omega\in\Omega:v(\omega)\in {\cal S}_{v_0}\})=\mathbb{Q}(\{\omega\in\Omega:v(\omega)\in {\cal W}_{S,v_0}\}).$

\noindent 
On the other hand, assume $\Hi_S$ to be invariant for the SME a.s. so that for all $\rho_0$ in ${\mathfrak I}_S(\Hi),$ and $t\geq 0$ 
\beqan\Pi_S\mathbb{E}(\rho_t)\Pi_S&=&\mathbb{E}(\Pi_S\rho_t\Pi_S)=%
\mathbb{E}(\rho_t)\\&=&\rho_0 + \int_0^t \left( {\cal F}( H,\rho_s) + \sum_{k=0}^r{\cal D}(L_k,\rho)
\right) ds.\eeqan Hence $\Hi_S$ is invariant for the semigroup dynamics driven by the ME.

\qed
\noindent It is also easy to see that subspace invariance can be checked by directly resorting to the simple deterministic equation \eqref{eq:lindet}.
\begin{corollary} \label{cor:invariance} A subspace $\Hi_S$ is invariant for the ME \eqref{eq:Lindblad} if and only if it is invariant for \eqref{eq:lindet} for any $u_t$.
\end{corollary}
\proof
From the proof of the previous proposition we have the ``only if'' implication.
On the other hand, if $\Hi_S$ is invariant for \eqref{eq:lindet}, it is easy to see that it must have the block structure of \eqref{eq:lindet1}, and hence $L_{k,Q}=0,\,\forall k,$ and $-iH_P-\um\sum_kL_{k,S}L_{k,P}^\dag=0.$

\qed
The next negative result will be a key part of the feedback-assisted stabilization strategies. Let us define the following nonnegative function
$
V_1(\rho):=\tr(\Pi_R\rho),
$
which is linear in the state, and is zero if and only if  the state is in ${\mathfrak I}_S(\Hi).$

\begin{proposition}\label{chidecrease} Assume $\Hi_R$ does not support invariant sets for the ME \eqref{eq:Lindblad}. Then  for every fixed time $T$
\beq\label{eq:chi} \chi(\rho) = \min_{t\in[0,T]}\mathbb{E}V_1(\rho_t)<1,\;\forall \rho_0\in\mathfrak{I}_R(\Hi).
\eeq
\end{proposition}
In order to prove this proposition, we will make use of the following property of linear deterministic autonomous systems.

\begin{lemma}\label{lemmadet} Consider a linear system $\dot x_t=Ax_t,\, x_t\in\Hi=\Hi_S\oplus\Hi_R.$ Assume $x_0\in\Hi_R,$ and that $\Hi_R$ does not contain any $A$-invariant subspace. Then for any $T\geq 0$ there exists $t\leq T$ such that $x_t\notin \Hi_R.$
\end{lemma}
\proof 
Assume by contradiction that for some $T$ and some $x_0\in\Hi_R$ we have $\Pi_S x_t =0,$ for all $0\leq t\leq T,$  with $\Pi_S$ the orthogonal projection on $\Hi_S.$
Hence $\Pi_S e^{At}x_0=0$ for $0\leq t \leq T,$ which is equivalent to say that $x_0$ is in the non-observable subspace for the LTI autonomous system 
\[\dot x_t = A x_t,\\y_t=\Pi_S x_t.\] This is true {\em iff} $0=\Pi_SA^kx_0$ with $k=0,\ldots,n-1.$  
Thus the subspace \[\Hi_{x_0}=\Span\{x_0,Ax_0,\ldots,A^{n-1}x_0\}\subseteq \Hi_R,\] and by the Caley-Hamilton theorem is $A$-invariant (and it is the smallest subspace that contains the whole trajectory), contradicting the initial assumption.
\qed

{\em Proof of Proposition \ref{chidecrease}.}
%
As in \eqref{eq:support_sub}, call ${\cal S}_{x_0}$ the closure of the set of trajectories of \eqref{eq:lindet} with piecewise-constant controls starting from $x_0,$ and call ${\cal W}^T_R$ the set of continuous paths that do not leave $\Hi_R$ up to time $T,$ which is closed in the uniform topology. Assume by contradiction that \eqref{eq:chi} is not true. Then it must be $\mathbb{Q}({\cal W}^T_R)=1$ (which is equivalent to $\mathbb{P}({\cal W}^T_R)=1$). But the smallest closed set of continuous paths that has probability 1 is ${\cal S}_{x_0},$ and hence it should be \[{\cal S}_{x_0}\subseteq {\cal W}^T_R.\]
However, if $\Hi_R$ does not support invariant sets for the ME, Corollary \ref{cor:invariance} implies that it does not contain invariant subspaces 
for the deterministic system \eqref{eq:lindet} as well. Then by Lemma \ref{lemmadet} every trajectory exits $\Hi_R$ in (any) finite interval in which the control is not zero, so ${\cal S}_{x_0}\nsubseteq {\cal W}^T_R,$ getting a contradiction. 
\qed
\section{Stabilization of the SME}
\subsection{Environment-Assisted Stabilization}\label{EAS}

Stabilization of the SME can be achieved in open loop, provided that the noise or the measurement operators can bridge the dynamics towards the target subspace $\Hi_S$, i.e. at least one of $L_{k,P}$ is nonzero. In
this case, it will suffice to: (i) prove that the if $\Hi_S$ is GAS for the ME, it is also GAS in probability for the corresponding SME;  and (ii) exploit Theorem \ref{theorem:ticozzi2009} to construct an open-loop Hamiltonian that renders $\Hi_S$ GAS for the ME.
The next proposition addresses the first issue.
\begin{proposition}
\label{prop:deterministic_to_stochastic}
If ${\mathfrak I}_S(\Hi)$ is GAS for
the ME (\ref{eq:Lindblad}) then it is GAS in probability for the SME (\ref{eq:SME-Ito1}).
\end{proposition}
\proof
Consider the function $V_1(\rho)$ as a
Lyapunov function candidate. It is zero iff $\rho_R=0$, i.e.,
$\rho\in{\mathfrak I}_S(\Hi)$, and positive for any
$\rho\notin{\mathfrak I}_S(\Hi)$. Moreover, direct computation yields
\[
{\cal A}V_1(\rho)=-\tr\left(\left(
\sum_k L_{k,P}^\dagger L_{k,P}
\right)\rho_R\right)
\]
where $ {\cal A} (\cdot) $ is the infinitesimal generator of the SME \eqref{eq:SME-Ito1}.
By the cyclic property of the trace, we observe that
${\cal A}V_1(\rho)\le 0$ for any $\rho\in{\mathfrak D}(\Hi)$. Thus, by Theorem
\ref{thm:stochastic_lyapunov}, it follows that
${\mathfrak I}_S(\Hi)$ is stable in probability.
To prove (by contradiction) it is also attractive, suppose that for some
$\rho_0\in{\mathfrak D}(\Hi)\setminus{\mathfrak I}_S(\Hi)$,
$\mathfrak{I}_S(\Hi)$ is not attractive for (\ref{eq:SME-Ito1}), i.e. with finite probability $p>0:$
\[
\mathbb{P}\left(\lim_{t\to\infty}\setdist{\Phi_t(\rho_0)}{{\mathfrak I}_S(\Hi)}=0\right)=1-p,
\]
where the probability measure $\mathbb{P}$ is the one induced by the Wiener process in \eqref{eq:SME-Ito1} and $ {\mathfrak d}( \cdot, \, \cdot) $ is defined in \eqref{eq:setdist}. Thus, by the properties of $V_1$, it must also be
\beq\label{limV1}
\mathbb{P}\left(\lim_{t\to\infty}V_1(\Phi_t(\rho_0))=0\right)=1-p.
\eeq
From the definition of the expectation,
\begin{eqnarray*}
\mathbb{E}\left[V_1(\Phi_t(\rho_0))\right]
=\int_{\{V_1(\Phi_t(\rho_0))>0\}} V_1(\Phi_t(\rho_0))d\mathbb{P}.
\end{eqnarray*}
Now, either the set $\{V_1(\Phi_t(\rho_0))>0\}$ has measure zero, contradicting \eqref{limV1}, or there exists a
(non-random) function $v(\rho_0)>0$ such that the following inequality
holds
\begin{eqnarray*}
\limsup_{t\to\infty}\mathbb{E}\left[V_1(\Phi_t(\rho_0))\right]
\\&=&\limsup_{t\to\infty}\int_{\{V_1(\Phi_t(\rho_0))>0\}} V_1(\Phi_t(\rho_0))d\mathbb{P}\\
&\ge& v(\rho_0)\limsup_{t\to\infty}\int_{\{V_1(\Phi_t(\rho_0))>0\}}d\mathbb{P}\\
&=&v(\rho_0)\limsup_{t\to\infty}\left\{
1-\mathbb{P}\left(V_1(\Phi_t(\rho_0))=0\right)
\right\}.
\end{eqnarray*}
By ``reverse'' Fatou's Lemma, we obtain:
\[
\limsup_{t\to\infty}E[V_1(\Phi_t(\rho_0))]\\
\ge v(\rho_0)
\left\{
1-\mathbb{P}\left(\limsup_{t\to\infty}V_1(\Phi_t(\rho_0))=0\right)
\right\}
=v(\rho_0) p.
\]
By linearity of the expectation and the trace
\[
\mathbb{E}[V_1(\Phi_t(\rho_0))]=V_1(\mathbb{E}[\Phi_t(\rho_0)]).
\]
Finally, by continuity of $V_1$, there exists a constant $k$ such that
\[
\limsup_{t\to\infty}\setdist{\mathbb{E}[\Phi_t(\rho_0)]}{{\mathfrak
I}_S(\Hi)}\ge\frac{v(\rho_0)p}{k}>0.
\]
Since the time evolution of $\mathbb{E}[\Phi_t(\rho_0)]$ is given by (\ref{eq:Lindblad}),
this would imply that ${\mathfrak I}_S(\Hi)$ is not attractive for
(\ref{eq:Lindblad}). Thus, the assertion is proved.
\qed
\begin{theorem}\label{thm:openloop}
Let $\Hi=\Hi_S\oplus\Hi_R$ and assume that at least one of $L_{k,P}$ is not zero, and that $L_{k,Q}=0,\,\forall k$. Then, there exists an
open-loop control Hamiltonian such that ${\mathfrak I}_S(\Hi)$ is
GAS in probability for the SME (\ref{eq:SME-Ito1}). 
\end{theorem}
\proof
Let $\tilde{H}$ be a Hamiltonian defined by
\[
\tilde{H}_o=
\left(\begin{array}{c|c}
0&\tilde{H}_{o,P}\\
\hline
\tilde{H}_{o,P}^\dagger&0
\end{array}\right)
\]
with $\tilde{H}_{o,P}=-H_{o,P}-\frac{i}{2}\left(\sum_k L_{k,S}^\dagger L_{k,P}\right).$  According to Proposition \ref{propinvariance}, the Hamiltonian $H'_o=H_o+\tilde{H}_o$ renders ${\mathfrak I}_S(\Hi)$ invariant for the ME. By Theorem~\ref{propiffinvariance} it is also a.s. invariant for (\ref{eq:SME-Ito1}). On the other hand, note that by the assumption on $L_{k,P}\neq 0$ for some $k$,
${\mathfrak I}_R(\Hi)$ is not invariant. Thus, by invoking Theorem
\ref{theorem:ticozzi2009}, we can find ${H}_c$ such that
${\mathfrak I}_S(\Hi)$ is attractive for the dynamics (\ref{eq:Lindblad})
with $H''_o={H}'_o+{H}_c$,
and still leaves the target subspace invariant.
Finally, by Proposition \ref{prop:deterministic_to_stochastic}, we have that
${\mathfrak I}_S(\Hi)$ is GAS in probability for the dynamics
(\ref{eq:SME-Ito1}) with $H''_o={H}'_o+{H}_c$.
\qed
{\noindent}
One can indeed prove that the hypothesis of Theorem \ref{thm:openloop} are also {\em necessary} for {open-loop} stabilization. This is formalized in the following Corollary, which completes Theorem \ref{thm:openloop}, establishing the limits of the open-loop approach. 

\begin{corollary}\label{cor:openloop}
Let $\Hi=\Hi_S\oplus\Hi_R$ and assume that $L_{k,P}=0$ for all $k$s. Then ${\mathfrak I}_S(\Hi)$ cannot be rendered
GAS in probability with open-loop control alone when the dynamics is given
by (\ref{eq:SME-Ito1}). 
\end{corollary}
\proof
If we chose $H_{o,P}=0,$ all the operators in \eqref{eq:SME-Ito1} would be block diagonal and it is easy to verify that also the set ${\mathfrak I}_R(\Hi)$ would be invariant for the resulting ME, and hence a.s. invariant for the SME dynamics. Hence ${\mathfrak I}_S(\Hi)$ could not be attractive in probability for \eqref{eq:SME-Ito1}. On the other hand, choosing $H_{o,P}\neq0$ would violate the conditions for the invariance of ${\mathfrak I}_S(\Hi)$ in Proposition \ref{propinvariance}.
\qed

\subsection{Feedback-Assisted Stabilization}\label{Sec:FAS}

\label{assumptions}
In the following, we will show that the addition of the time-dependent term, along with a suitable feedback law $u(\rho_t),$ will allow to render $\Hi_S$ GAS in expectation even when the open-loop method fails, namely when $L_{k,P}=0$ for any $k$. However, in order to do so, we will restrict our attention to a slightly less general class of problems and measurements. 

\noindent{\em Assumption 1:} Let $\dim(\Hi_S)=1,$ that is, we are seeking stabilization of a given pure state $\rho_d$.

The second assumption regards the structure of the feedback Hamiltonian. Recall that:
\beq\label{fham} H_t=H_o+u(t)H_f,\eeq
where the $u(t)$ is a control field and $H_f$ is {\em fixed}, while we assume as before we are free to design the time-independent part $H_o,$ in order to ``destabilize'' the undesired invariant sets for the dynamics. A trivial, necessary requirement for $H_f,$ in order to obtain the desired stabilization, is to dynamically ``connect'' the target (one-dimensional) subspace $\Hi_S$ to $\Hi_R.$

\begin{corollary}\label{cor:closedloop}
Let $\Hi=\Hi_S\oplus\Hi_R$ and assume that $L_{k,P}=0$ for all $k$'s. If the feedback Hamiltonian block $H_{f,P}=0,$ then there does not exist a choice of $H_o$ and $u(t)$ such that ${\mathfrak I}_S(\Hi)$
is GAS in probability for (\ref{eq:SME-Ito1}). 
\end{corollary}

\proof
The proof follows that of Corollary \ref{cor:openloop}: assuming $H_{f,P}=0,$ either we choose $H_{o,P}= 0,$ rendering ${\mathfrak I}_R(\Hi)$ invariant, or we choose $H_{o,P}\neq 0,$ rendering ${\mathfrak I}_S(\Hi)$ not invariant.\qed

\noindent It is then necessary to assume that the off-diagonal blocks of $H_f$ are not zero. On the other hand, in order to ensure invariance of ${\mathfrak I}_S(\Hi),$ by Proposition \ref{propinvariance} we need to assume a block diagonal open-loop Hamiltonian, that is $H_{o,P}=0.$ The following design assumption can be relaxed, but it would add up to the quite involved notation of the next sections.  

\noindent{\em Assumption 2:} In order to simplify the design and analysis of the control strategy, we assume $H_{f,S}=0$ and $H_{f,R}=0$.

\noindent Lastly, we restrict the class of measurements with a technical assumption that is required in the proof of Lemma \ref{convergence} below.

\noindent{\em Assumption 3:} $L_0$ is a normal operator with non-degenerate spectrum.

Since we already assumed that we are in the case where $L_{k,P}=0$ for all $k$s, and $L_{k,S}$ is a scalar, this last assumption affects only the block $L_{0,R}.$

\subsubsection{Construction of the open-loop control Hamiltonian}\label{Section:openloop}

The first step in our design approach, for both the open- and closed-loop strategies, is to design an open-loop Hamiltonian $H_o$ ensuring that there exist no invariant sets with support only in $\Hi_R.$ This result will be used in the open-loop approach directly (Section \ref{EAS}), and in the closed-loop approach to prove that $H'_o=H_o+\bar u H_f,$ for some constant control $\bar u,$ ``destabilizes'' the subspace $\Hi_R$ (Section \ref{FAS}). 

It will be convenient to consider a refinement of the Hilbert space decomposition by further splitting $\Hi_R$ into orthogonal subspaces. Any choice of orthonormal basis of each space in the right-hand side of
$\Hi=\Hi_S\oplus\Hi_C\oplus\Hi_Z$ induces a block
decomposition for matrices representing operators on $\Hi$:
\[
X=
\left(\begin{array}{c|c|c}
X_S&X_U&X_V\\
\hline
X_T&X_C&X_W\\
\hline
X_X&X_Y&X_Z
\end{array}\right),
\]
where $X_S,X_U,X_T,X_C$ are in fact scalar since $\Hi_S,\Hi_C$ have dimension one.
In particular, the choice of the $\Hi_C$ is made so the feedback Hamiltonian has a simple form: define $\Hi_C=\Span\{H_{f,P}^\dag\}\subset \Hi_R,$ where $H_{f,P}^\dag$ must non zero due to Corollary \ref{cor:closedloop}, and $\Hi_Z=\Hi_R\ominus\Hi_C.$ In the induced block decomposition, given the assumptions in Section \ref{assumptions} we then have:

\beq\label{hamform}
H_o=
\left(\begin{array}{c|c|c}
H_{o,S} & 0 & 0\\
\hline
0 & H_{o,C} & H_{o,W}\\
\hline
0 & H_{o,W}^\dag & H_{o,Z}
\end{array}\right), \;
H_f=
\left(\begin{array}{c|c|c}
0 & H_{f,U} & 0\\
\hline
H_{f,U}^* & 0 & 0\\
\hline
0 & 0 & 0
\end{array}\right), \eeq
\[
L_k=
\left(\begin{array}{c|c|c}
L_{k,S} & 0 & 0\\
\hline
0 & L_{k,C} & L_{k,W}\\
\hline
0 & L_{k,Y} & L_{k,Z}
\end{array}\right).
\]
The following technical Lemma will be used in constructing an open-loop Hamiltonian that prevents any subset of ${\mathfrak I}_Z(\Hi)$ from being invariant.
\begin{lemma}\label{aux} Given a ME \eqref{eq:Lindblad} and a decomposition $\Hi=\Hi_S\oplus\Hi_C\oplus\Hi_Z,$ the associate dynamics {\em does not admit} invariant subsets in ${\mathfrak I}_Z(\Hi)$ if for any $\rho\in{\mathfrak I}_Z(\Hi)$ either:
\beq \sum_k L_{k,W}\rho_Z L_{k,W}^\dag\neq 0,\label{Scond}\eeq
or 
\beq -iH_{o,W}\rho_Z +\sum_k \left(L_{k,W}\rho_Z L_{k,Z}^\dag-\smallfrac{1}{2}( L_{k,C}^\dag L_{k,W} + L_{k,Y}^\dag L_{k,Z})\rho_Z\right)\neq 0\label{Pcond}.\eeq

\end{lemma}

\proof By direct computation of the blocks of the ME, one finds its $C,W$-blocks to be
\beqa
&&{\cal L}_C=\sum_kL_{k,W}\rho_Z L_{k,W}^\dag,\nonumber\\ \label{condS}\\
&&{\cal L}_{k,W}=-iH_{o,W}\rho_Z +\sum_k\left(L_{k,W}\rho_Z L_{k,Z}^\dag -\smallfrac{1}{2}( L_{k,C}^\dag L_{k,W} + L_{k,Y}^\dag L_{k,Z})\rho_Z\right),\label{condP}
\eeqa
If a certain set ${\cal X}\in{\mathfrak I}_Z(\Hi)$ were invariant, then any state in ${\cal X}$ must satisfy ${\cal L}_C=0,$ ${\cal L}_W=0.$ If \eqref{Scond} or \eqref{Pcond} hold for any $\rho\in{\mathfrak I}_Z(\Hi),$ then ${\mathfrak I}_Z(\Hi)$ cannot be invariant.\qed

The following iterative procedure constructs an open-loop Hamiltonian term $H_c$ that ensures that no invariant set exists with support on $\Hi_Z$ alone.

\vspace{3mm}
{\em Design Procedure:} Pick a basis accordingly to the decomposition $\Hi_S\oplus\Hi_C\oplus\Hi_R,$ represent all the operators in matrix form and set the index $j=0$ to begin the procedure. Let us rename $\Hi_{C,Z}^{(0)}=\Hi_{C,Z},$ $H^{(0)}=H_o,$ $L_k^{(0)}=L_k.$ 
\begin{enumerate}
\item Consider $\Hi_C^{(j)}\oplus\Hi_Z^{(j)}$, and partition $H^{(j)},L_k^{(j)}$ accordingly;
\item If there exist $\bar k$ such that $L^{(j)}_{\bar k , W}\neq 0,$ build an orthogonal decomposition $\Hi_Z^{(j)}=\Hi_C^{(j+1)}\oplus\Hi_Z^{(j+1)},$ where $\Hi_C^{(j+1)}=\textrm{linspan}(L_{\bar k , W}^{(j)\dag})$ (notice that in matrix representation $L_{\bar k, W}^{(j)\dag}$ is a column vector of the same dimension of $\Hi_Z^{(j)}).$ In a basis chosen accordingly to the new decomposition, we have $L^{(j)}_{\bar k, W}=(\ell^{(j+1)},0,\ldots,0),$ with $\ell^{(j+1)}\neq 0$. 
By condition \eqref{Scond} of Lemma \ref{aux}, no state with support on $\Hi_S^{(j+1)}$ can be invariant.

\item If $L^{(j)}_{k,W} = 0$ for all $k$'s, proceed by computing $V_W^{(j)}:=-iH_{o,W}^{(j)}-\smallfrac{1}{2}\sum_k L_{k,Y}^{(j)\dag} L_{k,Z}^{(j)}.$ 
\begin{enumerate}
\item If $V_W^{(j)}=0$, add a Hamiltonian compensation term $H_c^{(j)}$ such that $H_{c,W}^{(j)}=(h^{(j)},0,\ldots,0)$ in the current basis and zero elsewhere, with $h^{(j)}>0$, and recalculate $V_P^{(j)}$;
\item Build an orthogonal decomposition $\Hi_Z^{(j)}=\Hi_c^{(j+1)}\oplus\Hi_Z^{(j+1)},$ where $\Hi_c^{(j+1)}=\textrm{linspan}(V_P^{(j)\dag}).$ By  condition \eqref{Pcond} of Lemma \ref{aux}, there cannot be invariant states with support in $\Hi_S^{(j+1)}.$
\end{enumerate}
\item If $\Hi_R^{(j+1)}=\{0\},$ exit the iteration. Otherwise, increment $j$ and proceed;
\end{enumerate}
The above procedure ends in a number of steps equal to the dimension of the initial $\Hi.$ Let us define:
\beq\label{eq:hamcorrection} H_c=\sum_jH_c^{(j)}.\eeq
\begin{proposition}\label{prop:openloop}
The dynamics driven by \eqref{eq:Lindblad}, with $L_k=L_k^{(0)},$ and $H'=H_o+H_c+H_f$ does not admit any invariant set with support in $\Hi_C\oplus\Hi_Z.$
\end{proposition}

\proof By following the procedure above, we construct a sequence $\Hi_C^{(j)},$ with $j=1,\ldots, m.$ Let us proceed by (finite) induction. Consider the last orthogonal subspace $\Hi_C^{(m)}$: By construction, it cannot support invariant states, and hence subset.  Next, assume by induction that no invariant state can have support on $\Hi_C^{(m-k)}\oplus\ldots\oplus\Hi_C^{(m)},$ and consider  adding $\Hi_C^{(m-k-1)}$. If a state has non-trivial support on $\Hi_C^{(m-k-1)},$ by the construction above it is not invariant. In fact either the upper-left or the off-diagonal block of the generator (opportunely partitioned) will be different from zero. 
Since $\Hi_Z^{(0)}=\Hi_C^{(1)}\oplus\ldots\oplus\Hi_C^{(m)},$ there cannot be invariant states with support on $\Hi_Z$ only. But given the form of $H_f$ and applying Proposition \ref{propinvariance}, there cannot be invariant states with support on $\Hi_C\oplus\Hi_Z$.
It is easy to show \cite{ticozzi2009} that, by linearity of the evolution and convexity arguments, if a certain set of states is invariant for the evolution, an invariant state must exist within the convex hull of the invariant set. 
Hence no invariant set with support on $\Hi_R$ exists.
\qed

\noindent{\bf Remark 1:} This result shows that there exists an open-loop Hamiltonian that ``destabilizes'' any state or subset with support in $\Hi_Z$ alone, and gives us a way to iteratively construct one. While the precise choice of the additional Hamiltonian we suggest may not be practical in most situations, a {\em generic} Hamiltonian of the form
\[H_c=
\left(\begin{array}{c|c|c}
0 & 0 & 0\\
\hline
0 & H_{c,C} & H_{c,W}\\
\hline
0 & H_{c,W}^\dag & H_{c,Z}
\end{array}\right) \]
will serve the scope. Proving this fact rigorously goes beyond the scope of the present work, but the argument would follow that of Section III.B of \cite{ticozzi-NV}. 

\noindent{\bf Remark 2:} Notice that by choosing the Hamiltonians as in \eqref{hamform} we decoupled the roles of the open-loop Hamiltonian and of the feedback Hamiltonian. In some other cases, assuming a different form of the feedback Hamiltonian $H_f$, would also serve the scope of destabilizing the undesired invariant sets once the control is set to some constant: this is the case, for example, of the spin-system stabilization of \cite{mirrahimi2007} which exploits the fact that the feedback Hamiltonian simultaneously couples all the energetic levels.

No matter how it is obtained, such an open-loop Hamiltonian is key element of our control strategy to globally stabilize $\rho_d.$

\subsubsection{Feedback stabilization of the SME}\label{FAS}

%
%
\noindent We are now ready to prove that the strategy developed for spin systems in \cite{mirrahimi2007} works also for the wider class of models we consider here, provided we allow for some freedom of choice in the open-loop Hamiltonian.
\begin{theorem}\label{thm:main}
Consider the SME \eqref{eq:SME-Ito1} with $H_t=H_o+H_c+u_tH_f,$ with $H_c$ as in \eqref{eq:hamcorrection}, and $L_0$ normal with non-degenerate spectrum. Set the control law to be:
\begin{enumerate}
\item If $\tr(\rho_{t}\rho_{d})\geq
\gamma$:
\beq u_{t}=-\text{tr}(i[H_f, \rho_{t}]\rho_{d})\label{controllaw};\eeq 
\item If $\tr(\rho_{t}\rho_{d})\leq
\gamma/2,$ $u_{t}=1$; 
\item If $\rho_{t} \in \mathcal{B}=\{\rho: \gamma/2 < \tr(\rho_{t}\rho_{d})<
\gamma\}$,
then $u_{t}=-\tr(i[H_f, \rho_{t}]\rho_{d})$ if $\rho_{t}$ last entered $\mathcal{B}$ through the boundary
$\tr(\rho_d\rho)=\gamma$, and $u_{t}=1$ otherwise.
\end{enumerate}
Then there exists a $\gamma >0$ such that $u_{t}$ globally
stabilizes \eqref{eq:SME-Ito1} around $\rho_{d}$ and
$\mathbb{E}(\rho_{t})\rightarrow \rho_{d}$ as $t \rightarrow \infty$.
\end{theorem}

We begin with a Lemma that ensure local asymptotic stability in probability of $\rho_d.$ Let us define ${\cal S}_{<\varepsilon}=\{\rho\in{\mathfrak{D}}(\Hi)|1-\tr(\rho\rho_d)<\varepsilon\}$. Notice that \eqref{controllaw} is used when $\rho\in {\cal S}_{<1-\gamma/2}.$

\begin{lemma}\label{convergence} The sample paths $\Phi(\rho,u)$ of \eqref{eq:SME-Ito1} that do not exit the set ${\cal S}_{<1-\gamma/2}$ converge in probability to $\rho_d$ as $t\rightarrow\infty.$
\end{lemma}
\proof
Consider $V_2(\rho)=1-\tr(\rho_d\rho)^2.$ 
It is straightforward to show that $V_2(\rho)\geq 0,$ with equality if and only if $\rho=\rho_d.$ Moreover, by using the fact that $[\rho_d,H_o]=[\rho_d,H_c]=[\rho_d,L_k]=0,$ we get:
\beqan {\cal A}V_2(\rho)&=&-2\tr\left(\rho_d({\cal F}(H_t,\rho))+\sum_k{\cal D}(L_k,\rho)\right)\tr(\rho_d\rho)\\&&-\tr(\rho_d{\cal G}(L_0,\rho))^2\\
&=&2u\tr(i[F_f,\rho]))\tr(\rho_d\rho)\\&&-4\eta\left(2\Re(L_{0,S})-\tr((L_0+L_0^\dag)\rho)\right)^2\tr(\rho_d\rho)^2.
\eeqan
Choosing $u=-\text{tr}(i[H_f, \rho_{t}]\rho_{d})$ we have that ${\cal A}V_2(\rho)\leq 0,$ and hence by Theorem \ref{thm:stochastic_lasalle}\footnote{The fact that all the conditions of the theorem are satisfied follows from Propositions 3.4 and 3.6 in \cite{mirrahimi2007}, that directly carry over to our setting.} we have that the paths that do not exit the set ${\cal S}_{<1-\gamma/2}$ converge in probability to the largest invariant set in ${\cal Z}=\{\rho\in{\mathfrak D}(\Hi)|{\cal A}V_2(\rho)=0\}\cap {\cal S}_{<1-\gamma/2}.$ Since the choice \eqref{controllaw} of control makes ${\cal A}V_2$ the sum of two squares, the points in ${\cal Z}$ must make zero both.
Focusing on the second term,  $\tr(\rho_d\rho_t)=0$ only outside the set ${\cal S}_{<1-\gamma/2},$ so we must have
\[2\Re(L_{0,S})-2\tr(L_0^H\rho)=0,\] $L_{0}^H$ being the Hermitian part of $L_0.$ Notice that $L_0^H$ must have the same block-diagonal structure of $L_0.$  Since
\beqan d\tr(L_0^H\rho)&=&-iu_t\tr\left(L_0^H({\cal F}(H_t,\rho)+\sum_k{\cal D}(L_k,\rho))\right)dt\\&&+2\sqrt{\eta}\tr(L_0^H{\cal G}(L_0,\rho))dW_t,\eeqan a necessary condition for the invariance is then
\beqan \tr(L_0^H{\cal G}(L_0,\rho))&=&\tr((L_0^HL_0+L_0^\dag L_0^H)\rho)-2\tr(L_0^H\rho)^2\\
&=& 2{\rm Var}(L_0^H,\rho)-\tr(i[L_0^H,L_0^S]\rho)=0,
\eeqan
where ${\rm Var}(X,\rho)=\tr(X^2,\rho)-\tr(X\rho)^2,$ and $L_0^S$ denotes the skew-Hermitian part of $L_0.$ Recall we assumed $L_0$ to be a normal matrix, so that $[L_0^H,L_0^S]=0.$ On the other hand
${\rm Var}(L_0^H,\rho)=0$ if and only $\rho=\Pi_j,$ with $\Pi_j$ a (rank-one) spectral projector of $L_0^H.$ Since the only such state in ${\cal S}_{<1-\gamma/2}$ is $\rho_d,$ we get the conclusion.
\qed

{\em Proof of Theorem \ref{thm:main}}:
The proof is identical to that of Theorem 4.2 of \cite{mirrahimi2007}, with only two differences:
\begin{itemize}
\item {\em Lemma 4.3 and 4.4 in \cite{mirrahimi2007} are substituted by the following argument:}\\ Proposition \ref{prop:openloop} ensures that there are no invariant subspaces for the ME \eqref{eq:Lindblad} when $H=H_o+H_c+H_f.$ Hence, Proposition \ref{chidecrease} guarantees that when $u_t=1,$ for any finite $T,$
\[\chi(\rho) = \min_{t\in[0,T]}\mathbb{E}V_1(\rho_t)<1;\] 
\item {\em Lemma 4.8 in \cite{mirrahimi2007} is substituted by Lemma \ref{convergence} above};
\end{itemize}
The rest of the Lemmi and arguments of the proof do not depend on the specific form of the operators $H,\{L_k\}$ but on the property of the solutions of \eqref{eq:SME-Ito1}, and carry over to our case directly.
\qed

\section{Conclusions and outlook}

In this paper we showed how open-loop Hamiltonian control can be sufficient for pure state (or subspace) stabilization when suitable measurement processes can be enacted, or dissipative environment are present. The key is to have non-hermitian noise operators, with an upper block-triangular matrix representation, that can effectively ``pump'' the state towards the desired subspace. In this case, stabilization of a pure state or a subspace for a time-invariant ME implies stabilization in probability for the underlying SME, linking the result to the analysis developed in \cite{ticozzi2008,ticozzi2009,ticozzi-NV}. 

When open-loop stabilization is not viable, we showed how adding a switching feedback controller on top of a suitable designed Hamiltonian perturbation, pure state stabilization can be achieved in expectation. The measurement operator is here requested to be associated to a normal matrix,  e.g. Hermitian or unitary. While the proof of stability in this second scenario relies on the use of the support theorem and on the results of \cite{mirrahimi2007}, it is worth highlighting some key technical differences and potential advantages of our result with respect to the existing ones. 

We assumed to have some freedom in constructing open-loop, time invariant Hamiltonians, and considered additional noise operators in addition to those induced by the measurement: it has been shown that, when the action of the noise does not prevent stabilization, it can actually be useful for the scope. For example, in the proposed design algorithm in Section \ref{Section:openloop}, the presence of extra noise, ``mixing'' the degrees of freedom in the $\Hi_R$ subspace could simplify the structure of the open-loop Hamiltonian, or even make it completely superfluous. 
In Proposition \ref{chidecrease}, and in particular in Lemma \ref{lemmadet}, we used a system-theoretic argument that allowed us to show that the switching controller destabilized other undesired equilibria for a quite large class of SMEs; similarly, Lemma \ref{convergence} shows that the control law \eqref{controllaw} locally stabilizes the target state for the same class of models. 
In the process, we derived also {\em necessary} condition on the structure of the Hamiltonians and the noise operators for achieving stabilization.

Lastly, in our approach the use of feedback is minimized, and the associated Hamiltonian has a simple structure. This suggests that the need for the computationally-intensive integration of the SME could be potentially reduced, addressing one of the critical obstacles to employing filtering-based feedback strategies for high-dimensional systems. More precisely, the switching times depend on the value of the function $f_1(\rho_t)=\tr(\rho_t\rho_d)=\rho_S\in \R$, and when the time-dependent stabilizing law is needed, this takes the form \[u_t=-\tr(i[H_f,\rho_t]\rho_d)=-2i\Im(H_{f,P}\rho_P^\dag)=-2i\Im(H_{f,U}\rho_{t,U}^*).\]
Hence, the real-time knowledge of only two parameters ($\rho_S,\rho_U$) of the state matrix is needed for the feedback law, hinting that a lower-dimensional estimator could be potentially used \cite{projectionfilter,filterreduction}. Further investigation of this intriguing possibility is needed in order to verify whether this is a viable research direction.

\appendix
\section{Stochastic differential equations: support theorem and Lyapunov stability}\label{stochstab}

In this appendix, we briefly recall the support theorem of \cite{mirrahimi2007} and a few tools of Lyapunov stability theory for stochastic systems, specialized to the case of stochastic master equations.

\subsection{Support theorem}
\begin{theorem}(\cite{mirrahimi2007}, Thm.~2.4)
\label{support}
  Let $M$ be a connected, paracompact $C^\infty$-manifold and let $X_k$,
  $k=0\ldots n$ be $C^\infty$ vector fields on $M$ such that all
  linear sums of $X_k$ are complete.
 Consider the Stratonovich equation
  \begin{equation}
      dx_t=X_0(x_t)\,dt+\sum_{k=1}^n X_k(x_t)\circ dW_t^k,
      \qquad x_0=x.
  \end{equation}
  Consider in addition the associated deterministic control system
  \begin{equation}
      \frac{d}{dt}x_t^u=X_0(x_t^u)+\sum_{k=1}^n X_k(x_t^u)u^k(t),
      \qquad x_0^u=x
  \end{equation}
  with $u^k\in\mathcal{U}$, the set of all piecewise constant
  functions from $\mathbb{R}_+$ to $\mathbb{R}$.  Then
  \begin{equation}
      \mathcal{S}_x=
      \overline{\{x^u_\cdot:u\in\mathcal{U}^n\}}\subset
      \mathcal{W}_x
\label{eq:support_sub}
  \end{equation}
  where $\mathcal{W}_x$ is the set of all continuous paths from
  $\mathbb{R}_+$ to $M$ starting at $x$, equipped with the topology
  of uniform convergence on compact sets, and $\mathcal{S}_x$ is the
  smallest closed subset of $\mathcal{W}_x$ such that
  $\mathbb{P}(\{\omega\in\Omega:x_\cdot(\omega)\in\mathcal{S}_x\})=1$.
\end{theorem}

Similar forms of this theorem can be found in \cite{kunita-flow,kunita-supp}.

\subsection{Stability for stochastic systems} 
Let ${\cal M}$ be an invariant set of (\ref{eq:SME-Ito1}), i.e., if
$\rho_0\in{\cal M}$, $\Phi_t(\rho_0)\in{\cal M}$ for all
$t\ge 0$, a.s.. Define a distance between a state and a set by
\begin{equation}
\setdist{\rho}{{\cal M}}:=\inf_{\sigma\in{\cal M}}\|\rho-\sigma\|.
\label{eq:setdist}
\end{equation}
Then, the stability of the invariant set is defined as follows.
\begin{definition}
An invariant set ${\cal M}$ of the SME (\ref{eq:SME-Ito1}) is said to
be
\begin{itemize}
\item stable in probability if for any $\varepsilon>0$
\begin{equation}
\label{set-stab-prob}
\lim_{\setdist{\rho_0}{{\cal M}}\to 0} P\left(
\sup_{0\le t<\infty}\setdist{\Phi_t(\rho_0)}{{\cal M}}\ge\varepsilon
\right)=0
\end{equation}
\item globally asymptotically stable in probability if it is stable in
probability and for any $\rho_0\in{\mathfrak D}(\Hi)$
\begin{equation}
P\left(\lim_{t\to\infty}
\setdist{\Phi_t(\rho_0)}{{\cal M}}=0
\right)=1.
\end{equation}
\end{itemize}
\end{definition}
The following two theorems are the stochastic counterparts of the
Lyapunov conditions and the LaSalle invariance principle
\cite{nishio2009,kushner1967,kushner1972,khasminskiy1980}. We denote by ${\cal A}$ the infinitesimal generator of the Markov semigroup associated to the SME
(\ref{eq:SME-Ito1}). Explicitly, it can be written in the following
``symmetrized'' fashion
\begin{equation}
\begin{split}
{\cal A} [\cdot] & = \frac{1}{2} \tr \left( ({\cal F}( H,\rho_t) + {\cal D} (
C, \rho_t )) \frac{\partial \, [\cdot] }{\partial \rho} + \frac{\partial
\, [\cdot] }{\partial \rho} ({\cal F}( H,\rho_t) + {\cal D} ( C, \rho_t ))
\right. \\ & \left. + \frac{1}{2} {\cal G} ^2(C, \rho_t) )
\frac{\partial^2 \, [\cdot] }{\partial \rho^2} + \frac{1}{2}
\frac{\partial^2 \, [\cdot] }{\partial \rho^2} {\cal G}^2 (C, \rho_t) )
\right).
\end{split}
\label{eq:inf-gen-Ito}
\end{equation}
We use the following
notations.
\begin{itemize}
\item~$V\in C({\mathfrak D}(\Hi),\R_+)$
\item~$Q_m:=\{\rho\in{\mathfrak D}(\Hi):V(\rho)<m\}$
\item $V$ is in the domain of the infinitesimal generator ${\cal A}$ of the diffusion.
\end{itemize}
\begin{theorem}
\label{thm:stochastic_lyapunov}
Assume that $\Phi_t(\rho_0)$ is a right continuous strong Markov process
defined at least until $\tau'>\tau_m$ with probability $1$, and that
there exists a function $V$ in the domain of ${\cal A}$ such that
\begin{enumerate}
\item~$V(\rho)=0$ for any $\rho\in{\cal M}$,
\item~$V(\rho)>0$ for any $\rho\notin{\cal M}$,
\item~${\cal A}V(\rho)\le 0$ for any $\rho\in Q_m$.
\end{enumerate}
Then ${\cal M}$ is stable in probability.
\end{theorem}
\begin{theorem}\label{thm:stochastic_lasalle}
Let ${\cal A}V(\rho)\le 0$ in $Q_m$. Suppose $Q_m$ has a compact
closure, $\Phi_{t}(\rho_0)$ is Feller continuous, and that
$P(\|\Phi_t(\rho_0)-\rho_0\|>\varepsilon)\to 0$ as $t\to 0$ for any
$\varepsilon>0$, uniformly for $\rho_0\in Q_m$. Then $\Phi_t(\rho_0)$
converges in probability to the largest invariant set contained in
$C_m=\{\rho\in Q_m:{\cal A}V(\rho)=0\}$ for almost all paths that do not leave $Q_m.$
\end{theorem}

\bibliographystyle{plain}


\begin{thebibliography}{1}


\bibitem{alicki-lendi}
R.~Alicki and K.~Lendi.
\newblock {\em Quantum Dynamical Semigroups and Applications}.
\newblock Springer-Verlag, Berlin, 1987.

\bibitem{altafini-open}
C.~Altafini.
\newblock Coherent control of open quantum dynamical systems.
\newblock {\em Phys. Rev. A}, 70(6):062321:1--8, 2004.

\bibitem{altafini2002}
C.~Altafini.
\newblock Controllability of quantum mechanical systems by root space
decomposition of su({N}).
\newblock {\em Journal of Mathematical Physics}, {\bf43}:2051--2062, 2002.

\bibitem{Arnold1}
L.~Arnold.
\newblock {\em Stochastic Differential Equations: Theory and Applications}.
\newblock Krieger Publishing Company, Malabar, Florida, 1974.

\bibitem{barchielli2009}
A.~Barchielli and M.~Gregoratti, \emph{Quantum Trajectories and Measurements in
 Continuous Time: The Diffusive Case}, ser. Lecture Notes in Physics,
 782.\hskip 1em plus 0.5em minus 0.4em\relax Springer, Berlin Heidelberg,
 2009.

\bibitem{marro-invariants}
G. Basile and G. Marro.
\newblock Controlled and Conditioned Invariants in Linear System Theory.
\newblock Prentice Hall, 1992.

\bibitem{belavkin1985}
V. P. Belavkin.
\newblock Nondemolition measurements and control in quantum dynamical systems.
\newblock In {\em Proceedings, Information Complexity and Control in Quantum
Physics, Udine 1985} (A. Blaquiere, S. Diner and G. Lochak Eds.).
\newblock pp.311-336. Springer-Verlag, Vienna-New York.

\bibitem{belavkin1992}
V. P. Belavkin.
\newblock Quantum stochastic calculus and quantum nonlinear
filtering.
\newblock Journal of Multivariate Analysis, 42:171-201,1992.

\bibitem{bouten2007}
L. Bouten, R. van Handel and M. R. James.
\newblock An introduction to quantum filtering,
\newblock SIAM J. Control Optim., {\bf 46}, 2199-2141, 2007.

\bibitem{petruccione}
H.~P. Breuer and F.~Petruccione.
\newblock {\em The Theory of Open Quantum Systems}.
\newblock Oxford University Press, UK, 2006.

\bibitem{brumer92}
P.~Brumer and M.~Shapiro, ``Laser control of molecular processes,''
 \emph{Annual Review of Physical Chemistry}, vol.~43, no.~1, pp. 257--282,
 1992.

\bibitem{wiseman-MIMO}
A. Chia, H. M. Wiseman, ``The Quantum Theory of MIMO Markovian Feedback with Diffusive Measurements,'' online preprint: http://arxiv.org/abs/1102.3098, 2011

\bibitem{wiseman-openloop}
J. Combes, H. M. Wiseman and A. J. Scott ``Replacing quantum feedback with open-loop control and quantum filtering,'' Phys. Rev. A 81, 020301(R), 2010.

\bibitem{doherty1999}
A.~C. Doherty and K.~Jacobs, ``Feedback control of quantum systems using
 continuous state estimation,'' \emph{Phys. Rev. A}, vol.~60, no.~4, pp.
 2700--2711, 1999.

\bibitem{doherty2000}
A.~C. Doherty, S.~Habib, K.~Jacobs, H.~Mabuchi, and S.~M. Tan.
\newblock Quantum feedback control and classical control theory.
\newblock {\em Phys. Rev. A}, 62:012105, 2000.

\bibitem{petersen-survey}
D.~Dong and I.~Petersen, ``Quantum control theory and applications: a survey,''
 \emph{IET Control Theory Appl.}, vol.~4, no.~12, p. 2651 -- 2671, 2010.

\bibitem{gorini1978}
V. Gorini, A. Frigerio, M. Verri, A. Kossakowski and E. C. G. Sudarshan.
\newblock Properties of quantum Markovian master equations,
\newblock Report on Mathematical Physics, {\bf 13}, 149-173, 1978.

\bibitem{ikeda-book}
N. Ikeda and S. Watanabe. 
\newblock Stochastic differential equations and diffusion processes, second ed. 
\newblock North-Holland, 1989.

\bibitem{khasminskiy1980}
R. Z. Khas'minskiy.
\newblock {\em Stochastic Stability of Differential Equations}.
\newblock Sijthoff and Noordhoff, Alphen aan den Rijn, The Netherlands, 1980.

\bibitem{yamamoto-delay}
K. Kashima, N. Yamamoto, Control of Quantum Systems Despite Feedback Delay,
IEEE Trans. Aut. Contr. 54(4):876-881, 2009

\bibitem{kunita-supp}
H.~Kunita.
\newblock Supports of diffusion processes and controllability problems.
\newblock In {\em Proc. Intern. Symp. SDE, Kyoto, 1976}, pages 163--185, 1978.

\bibitem{kunita-flow}
H.~Kunita.
\newblock {\em Stochastic flows and stochastic differential equations}.
\newblock Cambridge, 1990.


\bibitem{kushner1967}
H. J. Kushner.
\newblock Stochastic stability and Control,
\newblock Academic Press, 1967


\bibitem{kushner1972}
H.~J. Kushner.
\newblock Stochastic stability.
\newblock In R.~Curtain, editor, {\em Stability of stochastic dynamical
systems}, Lecture Notes in Math., Vol. 294, pages 97--124. Springer, Berlin,
1972.

\bibitem{lindblad1976}
G. Lindblad.
\newblock On the generators of quantum dynamical semigroups,
\newblock Commun. in Mathematical Physics, {\bf 48}, 119-130, 1976.


\bibitem{maassen-qp}
H.~Maassen.
\newblock Quantum probability applied to the damped harmonic oscillator.
\newblock In S.~Attal and J.M. Lindsay, editors, {\em Quantum Probability
Communications}, volume XII, pages 23--58. World Scientific, Singapore, 2003.

\bibitem{Mabuchi2005}
H.~Mabuchi and N.~Khaneja, ``Principles and applications of control in quantum
 systems,'' \emph{International Journal of Robust and Nonlinear Control},
 vol.~15, pp. 647--667, 2005.

\bibitem{nanomechanical1}
S.~Mancini, D.~Vitali, and P.~Tombesi, ``Optomechanical cooling of a
 macroscopic oscillator by homodyne feedback,'' \emph{Phys. Rev. Lett.},
 vol.~80, no.~4, pp. 688--691, Jan 1998.

\bibitem{meyer}
P.~A. Meyer.
\newblock {\em Quantum Probability for Probabilists}.
\newblock Lecture Notes in Mathematics: {1538}. Springer Verlag, 1995.

\bibitem{Mirrahimi2005}
M.~Mirrahimi, P.~Rouchon, and G.~Turinici.
\newblock Lyapunov control of bilinear {S}chr{\"o}dinger equations.
\newblock {\em Automatica}, 41:1987--1994, 2005.

\bibitem{mirrahimi2007}
M. Mirrahimi and R. van Handel.
\newblock Stabilizing feedback controls for quantum systems,
\newblock SIAM J. Control Optim., {\bf 46}, 445-467, 2007.

\bibitem{filterreduction}
A. E. B. Nielsen, A. S. Hopkins and H. Mabuchi, Quantum filter reduction for measurement-feedback control via unsupervised manifold learning, New J. Phys. (special issue on quantum control), 11, 105043, 2009.

\bibitem{nielsen2000}
M.~A. Nielsen and I.~L. Chuang.
\newblock {\em Quantum Computation and Quantum Information}.
\newblock Cambridge Univ. Press, 2000.


\bibitem{nishio2009}
K. Nishio, K. Kashima and J. Imura.
\newblock Global feedback stabilization of quantum noiseless subsystems,
\newblock 2009 American Control Conference, 1499-1504, 2009.

\bibitem{parthasarathy}
K.~R. Parthasarathy.
\newblock {\em An Introduction to Quantum Stochastic Calculus}, volume~85 of
{\em Monographs in Mathematics}.
\newblock Birkhauser, Basel, 1992.

\bibitem{Peirce1988}
A.~Peirce, M.~Dahleh, and H.~Rabitz, ``Optimal control of quantum mechanical
 systems: Existence, numerical approximations, and applications,'' \emph{Phys.
 Rev. A}, vol.~37, p. 4950, 1988.

\bibitem{shabani-lidar}
A.~Shabani and D.~A. Lidar.
\newblock Theory of initialization-free decoherence-free subspaces and
subsystems.
\newblock {\em Phys. Rev. A}, 72(4):042303:1--14, 2005.

\bibitem{orozco1}
W.~P. Smith, J.~E. Reiner, L.~A. Orozco, S.~Kuhr, and H.~M. Wiseman, ``Capture
 and release of a conditional state of a cavity \mbox{QED} system by quantum
 feedback,'' \emph{Phys. Rev. Lett.}, vol.~89, no.~13, p. 133601, 2002.

\bibitem{feedbackcooling}
D.~A. Steck, K.~Jacobs, H.~Mabuchi, S.~Habib, and T.~Bhattacharya, ``Feedback
 cooling of atomic motion in cavity qed,'' \emph{Phys. Rev. A}, vol.~74,
 no.~1, p. 012322, 2006.

\bibitem{Tannor1985}
D.~J. Tannor and S.~A. Rice, ``Control of selectivity of chemical reaction via
 control of wave packet evolution,'' \emph{J. Chem. Phys.}, vol.~83, pp.
 5013--5018, 1985.

\bibitem{ticozzi2008}
F. Ticozzi and L. Viola.
\newblock Quantum Markovian subsystems: Invariance, attractivity and
	control.
\newblock IEEE Trans. Automat. Control, {\bf 53}, 2048-2063, 2008.

\bibitem{ticozzi2009}
F. Ticozzi and L. Viola.
\newblock Analysis and synthesis of attractive quantum Markovian
	dynamics,
\newblock Automatica, {\bf 45}, 2002-2009, 2009.

\bibitem{ticozzi-NV}
F.~Ticozzi, R.~Lucchese, P.~Cappellaro, and L.~Viola.
\newblock Hamiltonian control of quantum dynamical semigroups: Stabilization
and convergence speed.
\newblock {\em submitted}, 2011. arxiv.org/pdf/1101.2452

\bibitem{tsumura2006}
K. Tsumura.
\newblock Global stabilization of $N$-dimensional quantim spin systems via continuous feedback.
\newblock Math. Eng. Tech. Rep. Univ. of Tokyo, 2006.


\bibitem{vanhandelstability}
R. van Handel, \newblock The stability of quantum Markov filters, \newblock Infin. Dimens. Anal. Quantum Probab. Relat. Top. 12, 153-172, 2009.

\bibitem{projectionfilter}
R. van Handel and H. Mabuchi,
\newblock Quantum projection filter for a highly nonlinear model in cavity QED, \newblock Journal of Optics B: Quantum and Semiclassical Optics, {\bf 7} (10), S226, 2005.

\bibitem{handel2005}
R. van Handel, J. K. Stockton and H. Mabuchi.
\newblock Feedback control of quantum state reduction,
\newblock IEEE Trans. Automat. Contr., {\bf 50}, 768-780, 2005.

\bibitem{wiseman-book}
H.~M. Wiseman and G.~J. Milburn.
\newblock {\em Quantum Measurement and Control}.
\newblock Cambridge University Press, 2009.


\bibitem{wiseman1993}
H. M. Wiseman and G. J. Milburn.
\newblock Quantum theory of optical feedback via homodyne detection.
\newblock Phys. Rev. Lett., {\bf 50}, 548-551, 1993.

\bibitem{wiseman1994}
H. M. Wiseman.
\newblock Quantum theory of continuous feedback.
\newblock Phys. Rev. A, {\bf 49}, 2133-2150, 1994.

\bibitem{wiseman2002}
H. M. Wiseman, S. Mancini and J. Wang.
\newblock Bayesian feedback versus Markovian feedback in a two-level
	atom.
\newblock Phys. Rev. A, {\bf 66}, 013807, 2002.

\bibitem{yamamoto2007}
N. Yamamoto, K. Tsumura and S. Hara.
\newblock Feedback control of quantum entanglement in a two-spin system,
\newblock Automatica, {\bf 43}, 981-992, 2007.
































\end{thebibliography}

\end{document}